\documentclass{appolb}
\usepackage{amsmath}
\usepackage{amssymb}
\usepackage{graphicx}
\begin{document}

\title{The logistic equation and a linear stochastic resonance}

\author{P.\ F.\ G\'ora\thanks{e-mail: gora@if.uj.edu.pl}
\address{M. Smoluchowski Institute of Physics and Complex Systems Research
Center, Jagellonian University, Reymonta 4, 30-059 Krakow, Poland}}

\maketitle

\begin{abstract}
We show analytically that a linear transmitter with correlated Gaussian 
white noises displays a stochastic resonance. We discuss the relation 
of this problem to a generalized noisy logistic equation.
\end{abstract}

\section{Introduction}

The logistic equation 

\begin{equation}\label{appb:logistic}
\dot x = ax(1-x)\,,
\end{equation}

\noindent $a>0$,
is one of the most frequently used, and undoubtedly most successful,
models of population dynamics. It is now one of the classical examples of
self-organization in many natural and artificial systems \cite{Eigen}. It is
virtually impossible to list its all applications --- for example, the Web 
search returns more than 60\thinspace000 links to pages and on-line 
publications containing the key words ``logistic equation''.

A natural generalization of the model \eqref{appb:logistic} is one
in which the deterministic growth rate, $a$, is perturbed by a stochastic
process:

\begin{equation}\label{appb:logistic-noise}
\dot x = (a+p\,\xi(t))x(1-x)\,,
\end{equation}

\noindent Here $\xi(t)$ is the stochastic term. Such equation has been
first discussed by Leung \cite{Leung} and later by many other authors.
In particular, authors of Refs.~\cite{McClintock1,McClintock2}, by Borel
summing a formal series for the expectation value $\left\langle x(t)\right\rangle$,
obtained expressions for the nonlinear relaxation time. The noise term
$\xi(t)$ was represented by the Gaussian white noise (GWN) in 
Refs.~\cite{Leung,McClintock1} and by a Gaussian colored noise in
Ref.~\cite{McClintock2}. Surprisingly, closed and mathematically exact 
expression for the
expectation value and the variance of $x(t)$ are still lacking;
we have not solved this problem, but we will present a~heuristics, based
on a simple transformation of Eq.~\eqref{appb:logistic-noise} that, as our
numerical simulations indicate, can be useful in predicting
the behavior of systems described by the noisy logistic equation.

Surprisingly, the problem of the noisy logistic equation is closely related 
to that of the linear stochastic resonance (LSR). The stochastic resonance
(SR) is an example of the constructive role of a noise, where the noise
and a dynamical system act together to reinforce a periodic signal
\cite{SR}; see also Ref.~\cite{SRreview} for a review. We should mention
that an aperiodic stochastic resonance has also been discovered \cite{ASR}.
The SR now seems to be ubiquitous and has been claimed ``an inherent
property of rate-modulated series of events'' \cite{bezrukov}.
There is, however, a long-standing debate whether the SR can be at all 
present in linear systems. This point is important because a~linear 
kinetics (relaxation) is the final stage of most dissipative
processes. It would be particularly important to establish the existence
of a LSR in systems driven by a GWN because such a noise represents the
standard equilibrium fluctuations of macroscopic physical systems.
The LSR has been first reported by Fuli\'nski in Ref.~\cite{Fulinski95}
an later discussed in Refs.~\cite{berdichevsky96,barzykin,berdichevsky99,JSP}.
However, it was soon pointed out that the LSR vanished in the GWN limit
\cite{berdichevsky96,berdichevsky99} or after averaging over the initial phase of
the signal \cite{barzykin}, or was restricted to transient times,
even though other constructive effects of the noise persisted in the
asymptotic regime as well \cite{JSP}. Short-lived or 
phase-dependent phenomena cannot be deemed ``unphysical'' --- on the
contrary, they may be very important in many situations, for example
in enzymatic reactions in living cells --- but there is a general
consensus that a fully-fledged
stochastic resonance should be characterized by a clear maximum of the
signal-to-noise ratio (SNR) which survives the phase averaging and persists
for long times, indicating the presence of an optimal noise level.

The first evidence for a LSR that displays a maximum of the SNR
was given only recently in Ref.~\cite{Cao}.
However, we have shown \cite{condmat-0304368} that this particular case of 
the SR was of a non-dynamical nature as the output of the dynamical
system discussed in Ref.~\cite{Cao} merely reproduced the spectral properties of the
input signal, and moreover, the resonant properties of the latter were
only an artifact of the chosen parameterization of the noises. In the
present paper we show analytically that a simple linear system with
correlated multiplicative and additive GWNs and a driving that consists
of a constant term and a periodic signal displays a maximum of the SNR
that lacks all the above-mentioned deficiencies: The maximum is present
in the asymptotic (long time) regime and not only for transient times, 
it survives averaging over the initial phase of the signal, and finally,
it is a result
of a~collective action of the system, the noises and the external driving
and not merely a reproduction of very particular properties of the 
input sequence.

\section{The logistic equation}

Assume that the noise term in the perturbed logistic equation 
\eqref{appb:logistic-noise} is a~GWN with $\left\langle\xi(t)\right\rangle=0$, 
$\left\langle\xi(t)\xi(t^\prime)\right\rangle=\delta(t-t^\prime)$. The equation
\eqref{appb:logistic-noise} can easily be solved for every specific
realization of the noise. The formal solution to Eq.~\eqref{appb:logistic-noise}
then reads

\begin{equation}\label{appb:logistic-noise-solution}
x(t) = \frac{1}{1+\frac{1-x_0}{x_0}\exp\left[-at
-p\,\int\limits_0^t\xi(t^\prime)dt^\prime\right]}\,.
\end{equation}

\noindent In most practical situations, one is interested not in specific
realizations of a stochastic process, but rather in moments of that process,
averaged over an ensemble. One is tempted, then, to calculate the moments directly
from Eq.~\eqref{appb:logistic-noise-solution}, expanding it in a power series:

\begin{eqnarray}\label{appb:rozbiezny}
\left\langle x(t)\right\rangle &=&
\left\langle\sum\limits_{n=0}^\infty
(-1)^n
\left(
\frac{1-x_0}{x_0}\exp\left[-at
-p\,\int\limits_0^t\xi(t^\prime)dt^\prime\right]
\right)^n
\right\rangle
\nonumber\\
&=&
\sum\limits_{n=0}^\infty(-1)^n
\left(\frac{1-x_0}{x_0}\right)^n e^{-nat}
\left\langle
\exp\left[-np\,\int\limits_0^t\xi(t^\prime)dt^\prime\right]
\right\rangle
\nonumber\\
&=&
\sum\limits_{n=0}^\infty(-1)^n
\left(\frac{1-x_0}{x_0}\right)^n e^{-nat}
\exp\left[\frac{1}{2}n^2p^2t\right],
\end{eqnarray}

\noindent where $\left\langle\cdots\right\rangle$ stands for an average
over the realizations of the noise $\xi(t)$. We have changed the order of
the summation and taking the average in Eq.~\eqref{appb:rozbiezny} and used
the well known fact that for a GWN 

\begin{equation}\label{appb:exp1}
\left\langle\exp\left[\int_{t_1}^{t_2}
f(t^\prime)\xi(t^\prime)dt^\prime\right]\right\rangle=
\exp\left[\frac{1}{2}\int_{t_1}^{t_2}\left(f(t^\prime)\right)^2dt^\prime\right].
\end{equation}

\noindent We can, however, see that the final series in Eq.~\eqref{appb:rozbiezny}
is clearly divergent and even though a series similar to it has been Borel summed 
and some useful information has been extracted from it \cite{McClintock1,McClintock2},
we conclude that the formal solution given by Eq.~\eqref{appb:logistic-noise-solution}
offers little insight into the properties of Eq.~\eqref{appb:logistic-noise}. 

Observe that a formal substitution

\begin{equation}\label{appb:substitution}
y = \frac{1}{x}
\end{equation}

\noindent leads immediately to 

\begin{equation}\label{appb:xinverse}
y(t) = 1+\frac{1-x_0}{x_0}\exp\left[-at
-p\,\int\limits_0^t\xi(t^\prime)dt^\prime\right]
\end{equation}

\noindent and if the process $y(t)$ is convergent (see below), 
after a sufficiently long time almost all realization of the
process do not differ from unity and the variance of $y(t)$
asymptotically vanishes. This observation does not constitute 
a~\textit{proof} that analogous statements can be made about
the logistic process $x(t)$, too, but provides a useful
heuristics about the long-time properties of the logistic
process. The substitution \eqref{appb:substitution}
gives also a possibility of a further generalization of the
noisy logistic equation that will be pursued in the next
Section.

\section{The linear transmitter}

With the substitution \eqref{appb:substitution} Eq.~\eqref{appb:logistic-noise}
changes into

\begin{equation}\label{appb:logistic-linearized}
\dot y=-(a+p\,\xi(t))y+(a+p\,\xi(t))\,.
\end{equation}

\noindent Note that this substitution does not affect the stable stationary 
point $x=1$ of Eq.~\eqref{appb:logistic} and its basin of attraction, but it
breaks for the unstable stationary point $x=0$ of Eq.~\eqref{appb:logistic}.
Therefore, the vicinity of the unstable stationary point may not be
represented correctly by the transformation \eqref{appb:substitution}.

The equation \eqref{appb:logistic-linearized} closely resembles the equation of
motion of a linear transmitter with multiplicative and additive noises
that has been the subject of our recent research \cite{PRE2001,condmat-0304055};
the principal difference is that the constant additive driving was lacking in
models discussed there. The equation~\eqref{appb:logistic-linearized} can
easily be generalized

\begin{equation}\label{appb:logistic-generalized}
\dot y = -(a+p\,\xi_m(t))y + (b + q\,\xi_a(t))\,,
\end{equation}

\noindent where both noises $\xi_m$, $\xi_a$ are GWNs, but they are correlated:
$\left\langle\xi_m(t)\xi_a(t^\prime)\right\rangle=c\,\delta(t-t^\prime)$.
We assume that the noises $\xi_m$, $\xi_a$ jointly form a two-dimensional
Gaussian process (this assumption simplifies the discussion but is not a~crucial
one~\cite{condmat-0304055}):

\begin{subequations}\label{appb:cholesky}
\begin{eqnarray}
\xi_m(t) &=& \xi(t)\,,\\
\xi_a(t) &=& c\,\xi(t) + \sqrt{1-c^2}\,\eta(t)\,,
\end{eqnarray}
\end{subequations}

\noindent where $\xi$, $\eta$ are uncorrelated (independent) GWNs and
$c\in[-1,1]$. We finally arrive at

\begin{equation}\label{appb:final}
\dot y = -(a+p\,\xi(t))y + b + qc\,\xi(t) + q\sqrt{1-c^2}\,\eta(t)\,.
\end{equation}

\subsection{Moments of the process $y(t)$}

A formal solution to Eq.~\eqref{appb:final} reads ($y(0)=0$)

\begin{equation}\label{appb:rozwiazanie}
y(t)=\int\limits_0^t e^{-a(t-t^\prime)}\exp\left[
-p\int\limits_{t^\prime}^t\xi(t^{\prime\prime})dt^{\prime\prime}\right]
\left(b + qc\,\xi(t^\prime) + q\sqrt{1-c^2}\,\eta(t^\prime)\right)dt^\prime\,.
\end{equation}

Unlike the solution \eqref{appb:logistic-noise-solution}, Eq.~\eqref{appb:rozwiazanie}
\textit{can} be used directly to calculate the moments:

\begin{eqnarray}\label{appb:srednia}
\left\langle y(t)\right\rangle &=&
b\int\limits_0^te^{-a(t-t^\prime)}
\left\langle\exp\left[
-p\int\limits_{t^\prime}^t\xi(t^{\prime\prime})dt^{\prime\prime}\right]
\right\rangle dt^\prime
\nonumber\\
&+&
q\sqrt{1-c^2}
\int\limits_0^te^{-a(t-t^\prime)}
\left\langle\eta(t^\prime)\exp\left[
-p\int\limits_{t^\prime}^t\xi(t^{\prime\prime})dt^{\prime\prime}\right]
\right\rangle dt^\prime
\nonumber\\
&+&
qc\int\limits_0^te^{-a(t-t^\prime)}
\left\langle\xi(t^\prime)\exp\left[
-p\int\limits_{t^\prime}^t\xi(t^{\prime\prime})dt^{\prime\prime}\right]
\right\rangle dt^\prime\,.
\end{eqnarray}

\noindent The first of the expectation values on the right hand side of 
Eq.~\eqref{appb:srednia} is of the form \eqref{appb:exp1}. The second vanishes
identically because the noises $\eta$ and $\xi$ are uncorrelated, and we have
calculated the last one in Ref.~\cite{condmat-0304055}:

\begin{equation}\label{appb:exp2}
\left\langle\xi(t^\prime)\exp\left[
-p\int\limits_{t^\prime}^t\xi(t^{\prime\prime})dt^{\prime\prime}\right]
\right\rangle
=
-\frac{1}{2}p\,\exp\left(\frac{1}{2}p^2(t-t^\prime)\right).
\end{equation}

\noindent After collecting all terms we get

\begin{equation}\label{appb:srednia2}
\left\langle y(t)\right\rangle
=\frac{b-\frac{1}{2}cpq}{a-\frac{1}{2}p^2}
\left(1-e^{-(a-\frac{1}{2}p^2)t}\right)
\mathop{\longrightarrow}\limits_{t\to\infty}
y_\infty=\frac{b-\frac{1}{2}cpq}{a-\frac{1}{2}p^2}\,.
\end{equation}

The limit in Eq.~\eqref{appb:srednia2} exists if $a-\frac{1}{2}p^2>0$;
otherwise the process is divergent. If we want to keep the interpretation
that $x=1/y$ is a measure of a population, the limit in Eq.~\eqref{appb:srednia2} 
makes a physical sense if $b-\frac{1}{2}cpq \geqslant0$.

We now calculate the variance of the process $y(t)$:

\begin{eqnarray}\label{appb:variance}
\left\langle y^2(t)\right\rangle 
&=&
\int\limits_0^t dt_1 \int\limits_0^t dt_2\, e^{-a(2t-t_1-t_2)} 
%\times{}
%\nonumber\\&&
\left\langle\exp\left[
-p\int\limits_{t_1}^t\xi(t^\prime)dt^\prime
-p\int\limits_{t_2}^t\xi(t^\prime)dt^\prime
\right]
\times{}
\right.
\nonumber\\&&
\left.
\left(b+qc\xi(t_1)+q\sqrt{1{-}c^2}\eta(t_1)\right)
\left(b+qc\xi(t_2)+q\sqrt{1{-}c^2}\eta(t_2)\right)
\vphantom{\exp\left[
-p\int\limits_{t_1}^t\xi(t^\prime)dt^\prime
-p\int\limits_{t_2}^t\xi(t^\prime)dt^\prime
\right]}
\right\rangle.
\nonumber\\
\end{eqnarray}

\noindent As we can see, one needs to know one more expectation value
that we have previously calculated in Ref.~\cite{condmat-0304055}:

\begin{gather}
\left\langle\exp\left[
\int\limits_0^T f(t^\prime)\xi(t^\prime)\,dt^\prime
\right]\xi(t_1)\xi(t_2)\right\rangle
={}\nonumber\\
\label{appb:exp3}
\left[\delta(t_1-t_2)+f(t_1)f(t_2)\right]
\exp\left[
\frac{1}{2}\int\limits_0^T \left[f(t^\prime)\right]^2\,dt^\prime
\right].
\end{gather}

\noindent It is now easy to verify that the variance \eqref{appb:variance}
exists if $a-p^2>0$. The full expression for the variance is rather long,
and therefore we present the final expression in the asymptotic regime
only:

\begin{gather}
D=\left\langle y^2(t)\right\rangle
-\left\langle y(t)\right\rangle^2 
\mathop{\longrightarrow}\limits_{t\to\infty}
\nonumber\\
\label{appb:D}
\frac{4b^2p^2-8abcpq+\left(4a^2-4a(1-c^2)p^2+(1-c^2)p^4\right)q^2}
{2(a-p^2)(p^2-2a)^2}\,.
\end{gather}

Let us consider the limiting cases. First, for the uncorrelated noises
we obtain

\begin{equation}\label{appb:D-c0}
D=\frac{b^2p^2+(a-\frac{1}{2}p^2)^2q^2}{2(a-p^2)(a-\frac{1}{2}p^2)^2}\,,
\quad c=0\,.
\end{equation}

\noindent As we can see, for the uncorrelated noises $D$ increases monotonically
both as a function of the amplitude of the multiplicative noise, $p$, and as 
a function of the amplitude of the additive noise, $q$. Next consider the case
of maximally correlated (or maximally anticorrelated) noises:

\begin{equation}\label{appb:D-c1}
D=\frac{(bp\mp aq)^2}{2(a-p^2)(a-\frac{1}{2}p^2)^2}\,,
\quad c=\pm1\,.
\end{equation}

\noindent The variance \eqref{appb:D-c1} can \textit{vanish} for certain values
of parameters: $D=0$ if $c=\pm1$ and $bp\mp aq=0$. For such parameters, almost all
realizations of the process $y(t)$ asymptotically remain equal their expectation
value $y_\infty$ despite the fact that the noises do not cease and
constantly act on the system. Note that in this case $y_\infty=\pm q/p\,$; the negative
value corresponds to an ``unphysical'' realization of the process (see the discussion
below Eq.~\eqref{appb:srednia2}).

In general, the variance $D$ exhibits, as a function of $q$, 
a minimum for all $\vert c\vert>0$ and 
appropriate values of other parameters; some of these sets of parameters may 
correspond to
``unphysical'' realizations. $D$ is, by construction, nonnegative and
equals zero only at its minimum for $c=\pm1$.

The fact that the linear transmitter described by Eq.~\eqref{appb:final} admits,
for the maximally (anti)correlated noises, realizations with a vanishing variance
can be shown quite elementarily: Let $c=\pm1$ and $bp\mp aq=0$. Then Eq.~\eqref{appb:final}
is equivalent to

\begin{eqnarray}
\label{appb:elem}
\dot y &=& -(a+p\,\xi(t))y + (b\pm q\,\xi(t)) = -(a+p\,\xi(t))\left(y\mp\frac{q}{p}\right),\\
\label{appb:elem-solution}
y(t)&=&\pm\frac{q}{p} + 
\left(y(0)\mp\frac{q}{p}\right)
\exp\left[-at -p\int\limits_0^t \xi(t^\prime)dt^\prime\right].
\end{eqnarray}

\noindent If the process is convergent, almost all realizations asymptotically approach
$y_\infty=\pm q/p$.

\subsection{Application to the logistic equation}

According to what we have said,
the linearized logistic equation \eqref{appb:logistic-linearized} clearly leads to
a solution with asymptotically vanishing $D$, or with $\left\langle y^2(t)\right\rangle
\!\!\to \left\langle y(t)\right\rangle^2 \to1$. While in general there is no simple relation
between the variances $D=\left\langle y^2(t)\right\rangle - \left\langle y(t)\right\rangle^2$
and
$D_x=\left\langle x^2(t)\right\rangle - \left\langle x(t)\right\rangle^2$, we expect
that if $D=0$ and $y_\infty\not=0$, $D_x$ vanishes, too. Observe that for sufficiently
large times, with $(a-\frac{1}{2}p^2)^{-1}$ being the characteristic time scale, almost
all realizations of $y(t)$ reach a constant value, and so does the inverse $1/y(t)=x(t)$.
We thus conclude that almost all realizations the process described by the logistic 
equation with a noisy growth rate \eqref{appb:logistic-noise} reach a constant value,
equal to the equilibrium value of the deterministic logistic equation \eqref{appb:logistic}.
Heuristically this means that once the process reaches a value $x(t)\simeq1$,
the right hand side of Eq.~\eqref{appb:logistic-noise} remains nearly equal to zero 
for all future times even though the growth rate fluctuates.
We have confirmed these results by direct numerical simulations.

\begin{figure}
\begin{center}
\vbox{\offinterlineskip
\includegraphics[scale=0.9]{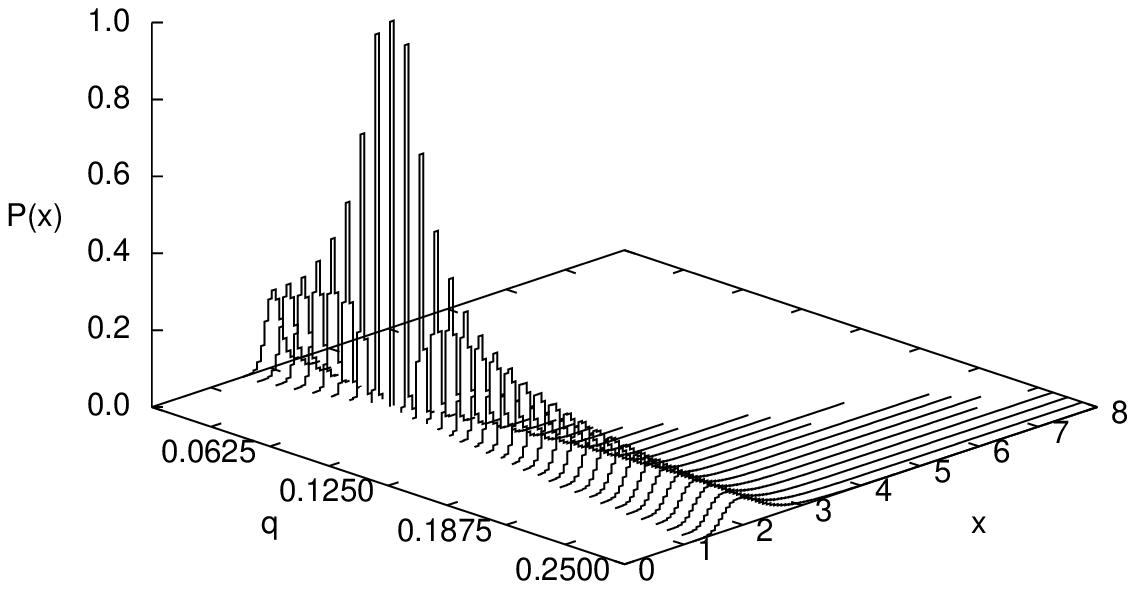}
\includegraphics[scale=0.9]{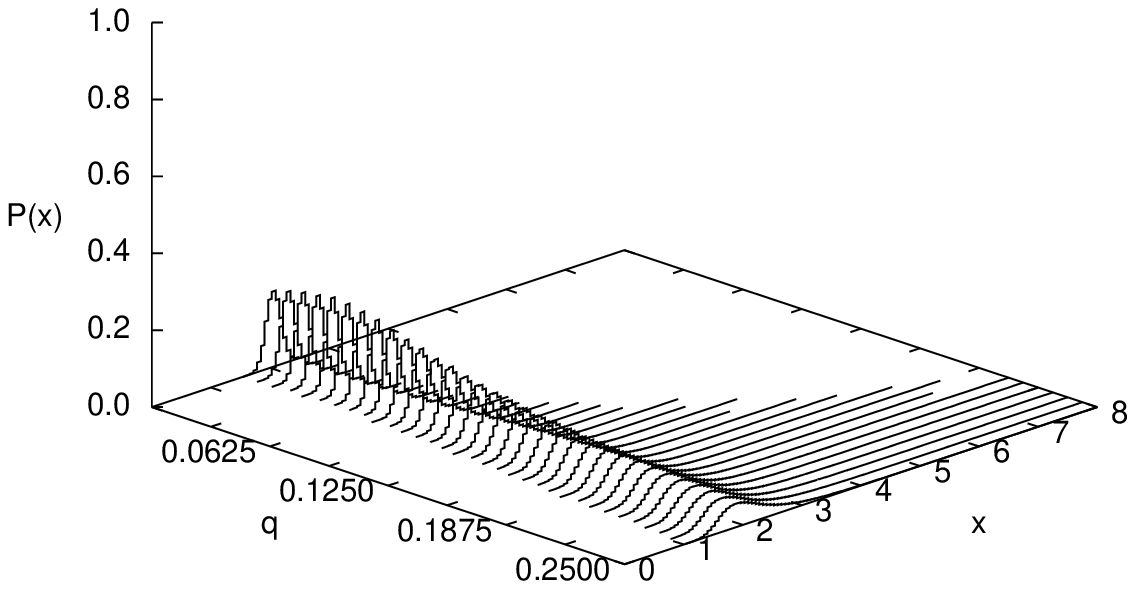}
}
\end{center}
\caption{Average distributions of values of $x(t)$ described by 
Eq.~\eqref{appb:logistic-noise-generalized} as a~function of the additive noise
level, $q$. Parameters are: $a=2$, $b=1$, $p=0.125$. The correlations are 
$c=1$ (upper panel) and $c=0.5$ (lower panel).}\label{appb:figure}
\end{figure}

The equation for the linear transmitter \eqref{appb:final} corresponds to a further
generalization of the logistic equation

\begin{equation}\label{appb:logistic-noise-generalized}
\dot x = - (a+p\,\xi(t))x + (b + qc\,\xi(t) + q\sqrt{1-c^2}\,\eta(t))x^2\,,
\end{equation}

\noindent where not only the growth rate, but also the limiting population
level fluctuate, and these fluctuations are correlated. The results of the previous
subsection suggest that in the presence of correlations, for certain values of the
parameters the variance of $x(t)$ should first shrink and then grow as the level
of the additive noise increases. We have performed numerical simulations to confirm this.
We have solved the nonlinear equation \eqref{appb:logistic-noise-generalized} 
using the Heun scheme \cite{Manella} with a time step $h=1/256$. The GWNs have
been generated by the Marsaglia algorithm \cite{Marsaglia} and the famous
Mersenne Twister \cite{Mersenne} has been used as the underlying uniform generator.
We have let the system to equilibrate for $2^{16}$ time steps, collected the results
for the next $2^{16}$ steps, calculated the distribution of the values of $x(t)$ 
that have appeared during the simulation, and averaged the results over 128 realizations
of the process. 

Selected results are presented in Fig.~\ref{appb:figure}. In case of maximal correlations
($c=1$, upper panel) the distribution of the values of $x(t)$ first gets narrower as 
$q$ increases, becomes practically $\delta$-shaped for $q=bp/a$, and then widens
as $q$ increases past the resonance. A similar, but much weaker, effect is observed
in case of partial correlations ($c=0.5$, lower panel). Note that for large
values of $q$ the distributions become skewed, with a marked preference
for values above the expectation value.

\section{A linear stochastic resonance}

So far we have discussed the systems without any external signal acting on them.
Now consider an additively coupled periodic signal:

\begin{equation}\label{appb:periodic}
\dot y = -(a+p\,\xi(t))y + b + qc\,\xi(t) + q\sqrt{1-c^2}\,\eta(t)
+ A\cos(\Omega t + \phi)\,,
\end{equation}

\noindent where $\phi$ is the initial phase of the signal, $A$ is the 
amplitude, $\Omega$ is the frequency, and all the other parameters are 
as above. We will, in addition to taking the usual average over 
realization of the noises, average over the initial phase, as otherwise
the correlation function would not correspond to a stationary series
\cite{Wiesenfeld}:

\begin{equation}\label{appb:doubleaverage}
\left\langle\left\langle y(t)y(t+\tau)\right\rangle\right\rangle
=
\frac{1}{2\pi}\int\limits_0^{2\pi}\left\langle y(t)y(t+\tau)\right\rangle d\phi\,.
\end{equation}

\noindent The calculations leading to the final formula for the correlation
function \eqref{appb:doubleaverage} are straightforward but tedious. The
expectation values \eqref{appb:exp1}, \eqref{appb:exp2}, and \eqref{appb:exp3}
are the most important ingredients. We obtain

\begin{gather}
\left\langle\left\langle y(t)y(t+\tau)\right\rangle\right\rangle
-
\left\langle\left\langle y(t)\right\rangle\right\rangle^2
\mathop{\longrightarrow}\limits_{t\to\infty}{}
\nonumber\\
\label{appb:correlation}
\frac{A^2\cos\Omega\tau}{2\left[(a-\frac{1}{2}p^2)^2+\Omega^2\right]}
+ \left[
\frac{A^2p^2}{4(a-p^2)\left[(a-\frac{1}{2}p^2)^2+\Omega^2\right]}
+ D \right]
e^{-(a-\frac{1}{2}p^2)\tau}\,,
\end{gather}

\noindent where $D$ is given by Eq.~\eqref{appb:D}. We can now calculate
the power spectrum of the process $y(t)$ as the Fourier transform of the
above correlation function (Wiener-Khinchin theorem) and then the 
signal-to-noise ratio (SNR) as the ratio of the power associated with the
signal, which corresponds to the Fourier transform of the first term 
on the right hand side of
Eq.~\eqref{appb:correlation}, to the power of the noisy background at
the frequency of the signal:

\begin{equation}\label{appb:SNR-def}
\mathrm{SNR} = 
10 \log_{10}\frac{P_{\text{signal}}}{P_{\text{noise}}(\omega=\Omega)}\,.
\end{equation}

\noindent By convenience, the SNR is usually measured in~dB.

\begin{figure}
\begin{center}
\includegraphics[scale=0.95]{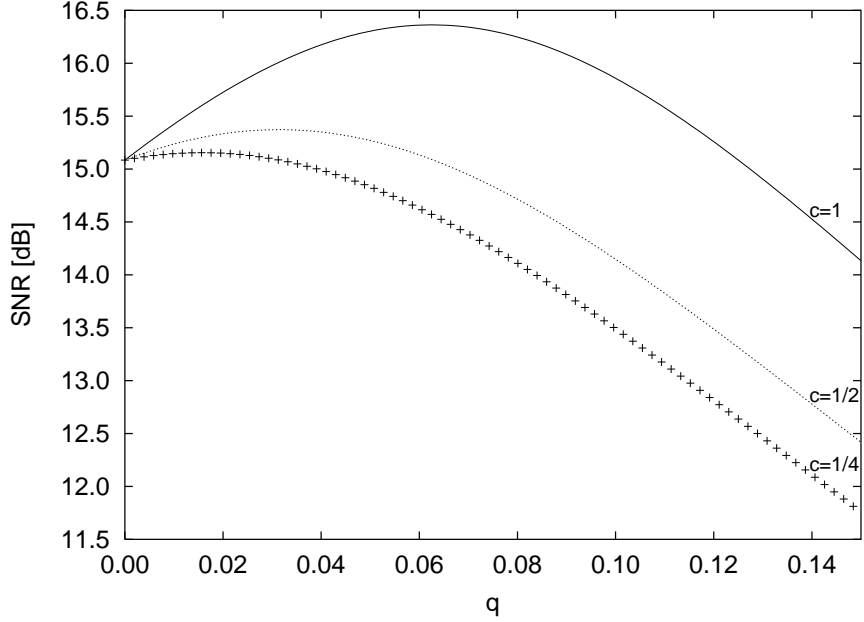}
\end{center}
\caption{Signal-to-noise ratio (SNR) for the linear transmitter \eqref{appb:periodic}
with various correlations between the noises: $c=1$ (solid line), $c=1/2$ (broken
line), $c=1/4$ (crosses). Amplitude of the signal $A=1$, frequency $\Omega=2\pi$.
Other parameters as on Fig.~\ref{appb:figure}. Curves with $c^\prime=-c$ and $b^\prime=-b$
are identical to those presented above, cf.\ Eq.~\eqref{appb:D}.}
\label{appb:psnr}
\end{figure}

For $c=1$ this procedure yields

\begin{equation}\label{appb:SNR}
\mathrm{SNR}=10\log_{10}
\frac{2A^2(a-p^2)(a-\frac{1}{2}p^2)\left[(a-\frac{1}{2}p^2)^2+\Omega^2\right]}
{A^2p^2(a-\frac{1}{2}p^2)^2
+2\left[(a-\frac{1}{2}p^2)^2+\Omega^2\right](bp-aq)^2}\,.
\end{equation}

\noindent As we can see, the SNR has, as a function of $q$, a clear maximum
for $bp-aq=0$. The presence of this maximum provides an unequivocal
evidence for the stochastic resonance.
It is easy to verify that for all $\vert c\vert>0$ and appropriate values
of other parameters the SNR has
a maximum as a function of the additive noise strength, but for small values
of $\vert c\vert$ this maximum is not very much pronounced, cf.\ 
Fig.~\ref{appb:psnr}. Note that also away from the maximum, larger values of
$\vert c\vert$ correspond to larger values of the SNR.

We have confirmed these results by direct numerical simulations. Numerical
results also indicate that there is a stochastic resonance in case of a
multiplicatively coupled signal. Analytical results are difficult to obtain
in this case due to the presence of transcendental function in the expression 
for the correlation function, cf.\ Refs.~\cite{PRE2001,condmat-0304055}.

\section{Discussion}

A clear and conclusive evidence that the linear system \eqref{appb:periodic}
displays a~fully-fledged stochastic resonance when driven by Gaussian white noises
is the principal result of this paper. The linear stochastic resonance reported 
here is characterized by a clear maximum of the SNR, persists for asymptotically long
times and survives averaging over the initial phase of the signal.
This result closes, we believe, the long debate whether the linear
stochastic resonance is at all possible. It is important to realize that 
there are \textit{two} factors that are
needed to produce the SR in the system \eqref{appb:periodic}: (i) the multiplicative
and additive noises have to be correlated, and (ii) there should be a constant
driving term, $b\not=0$, present. Berdichevsky and Gitterman in Ref.~\cite{berdichevsky99}
and the present author in Ref.~\cite{condmat-0304055} have also considered
correlated multiplicative and additive noises, but without the constant driving, no
stochastic resonance has been present, at least for the GWN case.
It is worth mentioning that the authors of Ref.~\cite{Cao} also discussed
a system with two correlated Gaussian noises, but in that case one of the
noises was coupled multiplicatively to the signal.

Note that the system \eqref{appb:periodic} is
linear in the sense that its equation of motion is linear with 
time-dependent coefficients, and thus we call the phenomenon
discussed above a ``linear stochastic resonance.'' There is, however, a point
in observation that since noise is meant to represent many unobserved and
unaccounted for degrees of freedom --- instead of considering impossibly
complicated microscopic motions, we mimic their effect by a reasonably
simple stochastic process --- a multiplicative (nonlinear) coupling between
the stochastic process and the observed degrees of freedom means a ``hidden''
nonlinearity. 

We have also shown that the problem of the linear transmitter is closely
related to the generalized noisy logistic equation. The analytical results
for the transmitter provide some heuristics that might be helpful in predicting
the behavior and properties of the noisy logistic process. However, the problem
of finding closed and mathematically exact formulas for the moments of the
latter remains open.

I am grateful to Dr.\ Ryszard Zygad\l{}o for helpful discussions and to
Prof.\ Andrzej Fuli\'nski for his constant encouragement.

\end{document}